\documentclass[%
 reprint,onecolumn,
 amsmath,amssymb,
 aps,
]{revtex4-2}

\usepackage{graphicx}
\usepackage{dcolumn}
\usepackage{bm}
\usepackage[utf8x]{inputenc}
\usepackage{float}
\usepackage{slashed}

\begin{document}


\title{Monte-Carlo Based QCD Sum Rules Analysis of $X_0(2900)$ and $X_1(2900)$}

\author{Halil Mutuk}
 \email{hmutuk@omu.edu.tr}
 \affiliation{ Department of Physics, Faculty of Arts and Sciences, Ondokuz Mayis University, 55139, Samsun, Turkey}

\begin{abstract}
Motivated by the very recent discovery of  fully open-flavor exotic states $X_0(2900)$ and $X_1(2900)$ by the LHCb Collaboration, we study possible interpretations of these exotic states by QCD Sum Rules method. We have calculated mass and decay constant by  traditional QCD Sum Rules analysis and Monte-Carlo based analysis. $X_0(2900)$ and   $X_1(2900)$ are studied in molecular and diquark-antidiquark tetraquark pictures, respectively. Obtained mass results agree well with the masses observed in the experiment. We also made a Monte-Carlo based analysis within QCD Sum Rules for the mass and decay constant distributions to investigate the possible assignments for the structures of $X_0(2900)$ and $X_1(2900)$.  
\end{abstract}

\pacs{Later}
\maketitle

\section{\label{sec:level1}Introduction}
The conventional quark model presents part of the properties of QCD in agreement with experimental data up to 2003 in which the first exotic hadron candidate $X(3872)$ was observed by Belle Collaboration \cite{Choi:2003ue}. According to quark model, hadrons can be classified into two groups: mesons made of quark-antiquark $(q\bar{q})$ pair and baryons made of three quarks, $(qqq)$. The spectrum of QCD is more richer than the spectrum of conventional quark model since any color neutral (colorless) configuration is possible in QCD. 

LHCb Collaboration announced an important and exciting discovery of exotic particles with open quark flavors in the  $D^-K^+$ of the $B^{\pm} \to D^+ D^- K^{\pm}$ channel \cite{Aaij:2020ypa}. The observed structure has been parametrized in terms of two Breit-Wigner resonances:
\begin{eqnarray}
X_0(2900) &:& J^P= 0^+, ~ M=2866 \pm 7 ~ \text{MeV}, ~ \Gamma_0 = 57 \pm 13 ~ \text{MeV}, \\
X_1(2900) &:& J^P= 1^-, ~ M=2904 \pm 5 ~ \text{MeV}, ~ \Gamma_0 = 110 \pm 12 ~ \text{MeV},
\end{eqnarray}
with global significance of more than 5$\sigma$. $X_0(2900)$ is a narrow state whereas $X_1(2900)$ is a broader one.  These states are $502 ~ \text{MeV}$ and $540 ~ \text{MeV}$ above than their corresponding $DK$ threshold, respectively. Both of these states decay into $D^- K^+$ and as a result their quark content should be $[ud\bar{s}\bar{c}]$. Since they have four different flavors, they could have an exotic nature. 

Actually, this is not the first  state consisting of four different flavor. D0 collaboration reported an exotic open quark flavor state $X(5568)$ decaying into $B_s^0 \pi$ in 2016 \cite{D0:2016mwd} but not confirmed by LHCb \cite{Aaij:2016iev}, CMS \cite{Sirunyan:2017ofq}, CDF \cite{Aaltonen:2017voc} and ATLAS\cite{Aaboud:2018hgx}. For recent studies on this exotic state see \cite{Sungu:2019ybf,Mutuk:2019uez}.  Based on this phenomena, confirmation of discovery of recent open quark flavor states by other collaborators may help understanding of the QCD, especially low-energy region of it. 

After the observation of the LHCb Collaboration, these two new states were studied by simple quark model \cite{Karliner:2020vsi}. In \cite{Hu:2020mxp}, the molecular nature of $X_0(2900)$ state is investigated. They extracted the mass positions of its heavy quark spin partners. Possible partners of $X_{0,1}(2900)$ were studied by two-body Coulomb and chromomagnetic interactions in \cite{He:2020jna}. In \cite{Liu:2020orv}, triangle singularity of $X_0(2900)$ and $X_1(2900)$ was studied in the $B^+ \to D^+ D^- K^+$ decay via the $\chi_{c1}K^{\ast +} D^{\ast -}$ and $D_{sJ}^+ \bar{D}_1^0 K^0$ rescattering diagrams. QCD Sum Rules method was applied in \cite{Zhang:2020oze} by using four possible interpolating currents with $J^P=0^+$. The mass spectra of open charm and bottom tetraquarks $qq\bar{q}\bar{Q}$ within an extended relativized quark model is calculated in \cite{Lu:2020qmp}. $D^{(\ast)}K^{(\ast)}$ system was studied within one-boson exchange model in molecular picture \cite{Liu:2020nil}. Molecular of compact tetraquark pictures of $X_0(2900)$ and $X_1(2900)$ were studied by QCD Sum Rules in \cite{Chen:2020aos}. They concluded $X_0(2900)$ as a molecule state and $X_1(2900)$ as compact diquark-antidiquark tetraquark state. In \cite{Wang:2020xyc}, axialvector-diquark-axialvector-diquark type and scalar-diquark-scalar-diquark type fully open flavor tetraquark states with $J^P=0^+$ were studied in QCD Sum Rules method. $X_0(2900)$ and $X_1(2900)$ states are studied whether two-body strong decays into $D^- K^+$ via triangle diagrams and three-body decays into $D^\ast \bar{K} \pi$ \cite{Huang:2020ptc}. In \cite{He:2020btl}, $X_0(2900)$ and $X_1(2900)$ are studied in qBSE approach. Tetraquarks composed of $ud\bar{s}\bar{c}$ are investigated with meson-meson and diquark-antidiquark structures in the quark delocalization color screening mode \cite{Xue:2020vtq}. Two states ($1^+$ and $2^+ $)stemming from the $D^\ast \bar{K}^\ast$ interaction was studied in \cite{Molina:2020hde}. In \cite{Burns:2020epm},   the LHCb vector $ud\bar{c}\bar{s}$ state X(2900) was studied whether it can be interpreted as a triangle cusp effect arising from $\bar{D}^\ast K^\ast$and $D_1K^(\ast)$ interactions. Mass and coupling of $X_0(2900$) are determined using the QCD two-point sum rule method in \cite{Agaev:2020nrc}. Branching ratios of $B^- \to D^- X_{0,1}(2900)$ was studied in \cite{Chen:2020eyu}. 	

An intriguing property of this experimental study is the widths of the resonances. The higher resonance $X_1(2900)$ with $J^P=1^-$ has a significantly larger width than the lower resonance $X_0(2900)$ with $J^P=0^+$. Considering the  $X_1(2900) \to D^-K^+$ decay is $P-$wave and $X_0(2900) \to D^-K^+$ is $S-$wave, this difference of widths may help to understand the possible interpretations of these resonances. For this purpose, we use QCD Sum Rules, an effective nonperturbative method to handle these resonances. 

The  paper is organized as follows. In Section \ref{sec:level2}, a brief introduction to QCD Sum Rules (QCDSR) and Monte-Carlo analysis are given. In Section \ref{sec:level3}, Monte-Carlo based analysis is done and numerical results are presented for $X_0(2900)$. Section \ref{sec:level4} is the same as previous section but for $X_1(2900)$. Section \ref{sec:level5} is a brief summary of this work.

\section{\label{sec:level2}QCD Sum Rules}

QCD Sum Rules are introduced by Shifman, Vainshtein and Zakharov in 1979 \cite{Shifman:1978bx} for mesons and generalized to baryons by Ioffe \cite{Ioffe:1981kw} in 1981. Due to the its relevance of QCD Lagrangian, it is one of the most used method in nonperturbative methods. In this model, hadrons are represented by suitably chosen interpolating currents made of quarks and in some cases together with gluons. Since these quarks are related to QCD Lagrangian, they are not constituent quarks which makes QCD Sum Rules model-independent since constituent quark masses change in different models. This model allows one to relate the hadron spectrum to the fundamental QCD Lagrangian. The fundamental idea behind the sum rules is to extrapolate correlation functions from large $q^2$ momentum transfer (short distances) where perturbative calculations can be done due to the asymptotic freedom, to low $q^2$ momentum transfer (large distances) where correlation functions are dominated by the resonances as a result of nonperturbative effects. Provided that there is some region where large $q^2$ and low $q^2$  representations overlap, one can compute physical properties of the resonances. 

In the first step, the correlation function is calculated taking into account short distances. This is the region where negative large momentum transfer occurs and therefore correlation function can be calculated by operator product expansion (OPE). In the second step, the correlation function is calculated at large distances where low momentum transfer occurs. In this region, correlation function can be evaluated by properties of hadron themselves. To match and overlap these two different correlation function which are written in two different kinematical region, spectral representation is used and hadronic properties can be extracted from this matching. 


A QCD Sum Rule study starts with writing the correlation function
\begin{equation}
\Pi(q^2)=i \int d^4x e^{iqx} \langle 0 | T[j(x)j^{\dagger}(0)]| 0\rangle. \label{corfunc}
\end{equation} 
Here $j(x)$ is the interpolating quark current, $q$ is the momentum of the related state, $T$ is the time ordering operator and $\vert 0 \rangle$ is the nonperturbative vacuum. Writing interpolating quark current is related to the quark content of the related hadron. As mentioned above, the correlation function in Eq. \ref{corfunc} can be calculated in two different ways: the first way includes evaluation of correlation function by using OPE and the second way includes a phenomenological usage in terms of related hadron parameters.

\subsection{The OPE Side}
Quark and gluon fields are the degrees of freedom of QCD, and we have to take care the populating of these fields into the QCD vacuum which is the complex structure of QCD. We can work analytically in the perturbative regime so that we can calculate perturbative part of $\Pi(q)$ in Eq. (\ref{corfunc}). Due to the complex structure of QCD vacuum, expectation values of the operators of quarks fields and gluon fields are non-zero resulting what we call condensates. Condensate (vacuum field) contributions to the correlation function can be calculated via Wilson OPE. 

Applying OPE to the Eq. (\ref{corfunc}) gives
\begin{equation}
\Pi(q^2)=i \int d^4x e^{iqx} \langle 0 | T[j(x)j^{\dagger}(0)]| 0\rangle= \sum_d C_d(Q^2)\left\langle  O_d \right\rangle, \label{ope}
\end{equation} 
where $C_d(Q^2)(Q^2=-q^2)$ are the Wilson's coefficients, and $\left\langle  O_d \right\rangle$ is the expectation value of the local operators. The coefficients include only short distance effects and can be calculated perturbatively. Long-distance effects which are non-perturbative are contained in the local operators. In the right hand side of Eq. (\ref{ope}), the operators are ordered by their dimensions $d$.  

In the OPE side of correlation function, one evaluate it in terms of the OPE expansion. Since this expansion is simply an expansion of the operators contributed from QCD condensates, it must convergence in order to obtain physical results. Using dispersion relation, the OPE side of Eq. (\ref{corfunc}) can be represented as
\begin{equation}
\Pi^{\text{OPE}}(q^2)=\int_{s_{min}}^\infty \frac{\rho^{\text{OPE}}(s)ds}{s-q^2},
\end{equation} 
where
\begin{equation}
\rho^{\text{OPE}}(s)=\frac{1}{\pi}\text{Im}[\Pi^\text{OPE}(s)],
\end{equation}
is the spectral density function and $s_{min}$ is a kinematical limit. This kinematical limit can be viewed as an adequate energy to create studying hadron. 

\subsection{The Phenomenological Side}
In previous subsection, QCD side of the correlation function was evaluated by the QCD degrees of freedom, which are quark and gluon fields. In the phenomenological side, hadrons themselves are being used as degree of freedom. In order to do this, intermediate states of the studying hadron are inserted in the correlation function. The correlation can be written as follows:
\begin{equation}
\Pi^{\text{Phen}}(q^2)=\frac{\langle 0 \vert j \vert h(q) \rangle \langle h(q) \vert j^\dagger \vert 0 \rangle}{q^2-m^2} + \int_{s_0}^\infty ds \frac{\rho(s)}{s-q^2} + \mbox{subtraction terms}, \label{phen}
\end{equation} 
where $\rho(s)$ is the spectral density of higher states and continuum. In order to QCDSR be useful, one must parametrize spectral density with a small number of parameters. The lowest resonance is often fairly narrow, where as higher-mass resonances are broader. Therefore, one can parametrize the spectral function as a single sharp pole decoding the lowest resonance of mass $m$, plus a smooth continuum representing higher mass resonances as
\begin{equation}
\rho(s)=\lambda^2 \delta (s-m^2)+ \rho^{\text{cont}}(s),
\end{equation}
where $\lambda$ gives the coupling of the current to the lowest mass hadron $m$ as $\langle 0 \vert j \vert H \rangle=\lambda$. Putting this relation into the Eq. (\ref{phen}), one can get the following expression for the phenomenological side: 
\begin{equation}
\Pi^{\text{Phen}}(q^2)=\frac{\lambda^2}{q^2-m^2}+ \int_{s_0}^\infty \frac{ds \rho^{\text{OPE}}(s) }{s-q^2} + \mbox{subtraction terms}.
\end{equation}
Subtraction terms are suppressed when Borel transform is applied:
\begin{equation}
\mathcal{B}_M^2 [\Pi(q^2)]=\lim_{\substack {-q^2, n \to \infty \\ -q^2/n=M^2}} \frac{(-q^2)^{n+1}}{n!} \left( \frac{d}{dq^2} \right)^n \Pi(q^2).
\end{equation} 
Borel transformation kills the subtraction terms. It also kills or better to say suppress contributions of excited resonances and continuum in the phenomenological side. Furthermore, in the OPE side, it suppresses factorially the contribution from higher dimension condensates which have inverse power of $q^2$. After Borel transformation, one can able to extract the decay constant $\lambda$ and mass $m$ from the sum rule. The sum rule can be written as

\begin{equation}
\lambda^2 e^{-\frac{m_H^2}{M^2}}=\int_{s_{min}}^{s_0} ds \rho^{\text{OPE}}(s) e^{-\frac{s}{M^2}} \label{sumrule}
\end{equation}
By taking the derivative of Eq. (\ref{sumrule}) with respect to $1/M^2$ and dividing the result by Eq. (\ref{sumrule}), one can obtain the mass
\begin{equation}
m_H^2=\frac{\int_{s_{min}}^{s_0} ds s \rho^{\text{OPE}}(s) e^{-s/M^2}}{\int_{s_{min}}^{s_0} ds \rho^{\text{OPE}}(s) e^{-s/M^2}},
\end{equation}
and decay constant $\lambda$ as
\begin{equation}
\lambda^2 m_H^2 e^{-\frac{m_H^2}{M^2}}=\int_{s_{min}}^{s_0} ds \rho^{\text{OPE}}(s) e^{-\frac{s}{M^2}}.
\end{equation}
For more details of the QCDSR technique see \cite{Nielsen:2009uh,Albuquerque:2018jkn}.

\subsection{Monte-Carlo Analysis} 
In the standard QCDSR analysis,  the spectral density of the higher states and the continuum are parameterized by using quark-hadron duality which assumes $\rho(s)=\rho^{\text{QCD}}(s)$ for $s>s_0$. This means that higher states and continuum contributions to the spectral density is approximated by the spectral density which is written in terms of QCD parameters and this parameters are sources of uncertainties for the method.

The major source of uncertainty resulting in the QCDSR technique is the vacuum condensates. Nonperturbative quantum fluctuations generate condensates and these condensates emerge in the OPE side of the correlation function. Since it is possible to study higher dimension operators in the OPE side of the correlation function, one needs to assume factorization principle to replace vacuum expectation values of higher dimensional operator by products of lower dimensional operators. This assumption become problematic in the calculations. Due to the lack of complete understanding of QCD, many features of these condensates are not yet well understood  \cite{Gubler:2018ctz}. (Although it is beyond the scope of this work, it was claimed by Brodsky et al. in \cite{Brodsky:2010xf} that QCD condensates are properties  of hadrons themselves. This conclusion was challenged in \cite{Reinhardt:2012xs}. Brodsky et al. in another work suggest that confinement is related to condensates and conclude that if quark-hadron duality is a fact in QCD, then  condensates are contained within hadrons \cite{Brodsky:2012ku}.)

The other sources of uncertainties of the QCDSR approach are unfactorized condensate values, factorization of higher dimensional operators, $\alpha_s$ corrections, truncation of the OPE, and choosing of the working regions for matching the QCD and phenomenological sides of the obtained sum rules. The conventional QCDSR analysis include the following steps:
\begin{itemize}
\item choosing appropriate values for the Borel parameters which give good results generally with respect to experimental observables. A working region of Borel parameters should be selected  in order to obtain the stability and reliability of the results.
\item a selection of the Borel window without giving a careful attention to OPE convergence nor maintaining ground state dominance of the phenomenological side of the sum rules.
\item fixing of input parameters (such as the continuum threshold) to preferred values. 
\item claiming an accuracy for QCDSR predictions without supporting calculations. Generally it is seeked a mild dependence of the results with respect to model parameters. This is called stability analysis.  In this analysis, fit parameters are monitored as a single condensate value is varied by a specific amount.
\end{itemize}
This traditional QCDSR analysis can be found in any paper of QCDSR studies. The sum rules could be analyzed in a certain range of the Borel parameter $M^2$ (Borel window) and continuum threshold $s_0$, in which the OPE series reaches convergences and the continuum contribution is suppressed. The OPE convergence is ensured with two criteria:  contributions of the highest dimensional operators in the OPE side should not be too large (geenrally less than 10 \% of the total OPE contributions). This give an upper bound of Borel parameter. On the other side, the continuum contributions should not be larger than the pole contributions. Excited state contributions over total contribution  should be less than 50 \% which gives the lower bound for Borel parameter. One important point here is that excited state contributions depend on continuum threshold and this value will be determined in QCDSR analysis. Therefore, we cannot determine lower bound of Borel parameter before carrying out QCDSR analysis. The common attitude is to make an initial guess for lower bound of Borel parameter. If the extracted physical properties are not sensitive to the variation in the Borel window with respect to different continuum threshold values, reliable predictions can be obtained by demanding the OPE convergence. This was done for exotic states in  \cite{Narison:2014vka,Turkan:2017pil,Mutuk:2018zxs}. 

Apart from systematic uncertainties, QCDSR method is a powerful technique among nonperturbative methods since it is relativistic, an analytic method and  related to the QCD Lagrangian. The main purpose of QCDSR technique is to extract physical properties, mostly  resonance mass and coupling constant. As mentioned above, especially fixing $s_0$ by hand, systematic uncertainties related to the calculation of OPE side and determination of phenomenological quantities with uncertainties related to the sum rule calculations are addressed  and an alternative method is presented by a Monte-Carlo analysis by Leinweber in \cite{Leinweber:1995fn}.

Leinweber introduced Monte-Carlo based uncertainty analysis for QCDSR predictions. For some applications of this method see \cite{Zhang:2013rya,Huang:2014hya,Wang:2016sdt,Kucukarslan:2015urd}. To do this, we first need to estimate the standard deviation $\sigma_{\text{OPE}}(M^2)$ of $\Pi^{\text{OPE}}(M^2)$ at any point in the sum rule interval. In this work, this estimation can be done by randomly generating 250 set of Gaussian distributed input parameters of QCD (condensates and $\Lambda_{\text{QCD}}$) with given uncertainties. Once the standard deviation is obtained the phenomenological output parameters ($s_0$, $\lambda$, $m$) can be obtained by minimizing a weighted $\chi^2$:
\begin{equation}
\chi^2=\sum_{j=1}^{n_B} \frac{(\Pi^{\text{OPE}}(M^2)-\Pi^{\text{Phys}}(M^2,s_0,\lambda,m))^2}{\sigma_{\text{OPE}}(M^2)},
\end{equation}
where $M^2=M^2_{min}+(M^2_{max}-M^2_{min})(j-1)/(n_B-1)$ which means dividing the sum rule window into $(n_B-1)$ even parts. After obtaining $\sigma_{\text{OPE}}(M^2)$, 3000 sets of Gaussian distributed input parameters with the same given uncertainties will
be generated. One can minimize $\chi^2$ to extract a set of phenomenological
output parameters for each set and the uncertainties of continuum threshold $s_0$, decay constant $\lambda$, and mass $m$. Since the distribution of input parameters occurs randomly we can impose an important constraint: results obtained with the $s_0 < m^2$ condition will be excluded since the original sum rules of QCD is formulated with the ansatz $s_0>m^2$. 

In the next two sections, we will make a traditional analysis vs. Monte-Carlo analysis comparison for the $X_0(2900)$ and $X_1(2900)$ resonances.

\section{\label{sec:level3}Traditional and Monte-Carlo Analysis of $X_0(2900)$ with $J^P=0^+$}
The observed peak of $X_0(2900)$ by LHCb is around 2.9 GeV. This is very close to the $D^{\ast -}K^{\ast +}$ threshold at 2902 MeV. In this section, we assume $X_0(2900)$ has a molecular picture and created in the vacuum by the following interpolating current with $J^P=0^+$:
\begin{equation}
j(x)=\left[ \bar{c}^a(x) \gamma_5 d^a(x) \right] \left[ \bar{s}^b(x) \gamma_5 u^b(x) \right].
\end{equation}
Here the subscripts $a$ and $b$ are color indices and, $u$, $d$, $s$ and $c$ represent the $up$, $down$, $strange$ and $charm$ quark fields, respectively. The components $\bar{c}_a \gamma_5 d_a$ and $\bar{s_b} \gamma_5 u_b$ in the interpolating current are two generic meson operators. They couple to the $D^{\ast -} $ and $K^{\ast +}$ mesons, respectively. The current can couple to the $S$ wave $D^{\ast -}K^{\ast +}$ molecular state with $J^P=0^+$. The coupling of current $j(x)$ to the $X_0$ state can be defined as
\begin{equation}
\left\langle 0 \vert j \vert X_0 \right\rangle = \lambda_{X_0}.
\end{equation}

The QCDSR calculation starts with obtaining correlation function in terms of the physical degrees of freedom. This is the first step and ends up with Borel transformed form of the function $\Pi^{\text{Phys}}(q)$:
\begin{equation}
\mathcal{B}_{q^2} \Pi^{\text{Phys}}(q)=m_{X_0}^2 \lambda_{X_0}^2 e^{-m_{{X_0}}^2/M^2} + \cdots .
\end{equation}
The next step is to find the theoretical expression for the same function, $\Pi^{\text{OPE}}(q)$. Contracting the quark fields yields
\begin{eqnarray}
\Pi^{\text{OPE}}(q)= i \int d^4 x e^{iqx} \text{Tr} [\gamma_5 S_c^{aa^\prime}(x)S_d^{aa^\prime}(-x)] \text{Tr} [\gamma_5 S_u^{aa^\prime}(x)S_s^{aa^\prime}(-x)],
\end{eqnarray}
where $S_q^{ab}(x)$ ($q=u,d,s$) and $S_c^{ab}(x)$ are the light and heavy quark propagators, respectively. The spectral densities are given in the appendix. The light quark propagator $S_q^{ab}(x)$ reads as
\begin{eqnarray}
S^{ab}_q(x)&=&i \delta_{ab} \frac{\slashed x}{2\pi^2 x^4}- \delta_{ab} \frac{\langle \bar{q}q \rangle}{12} + i\delta_{ab} \frac{\slashed x m_q \langle \bar{q}q \rangle}{48}  
 - \delta_{ab} \frac{x^2}{192} \langle \bar{q} g_s \sigma G q \rangle + i \delta_{ab} \frac{x^2 \slashed x m_q }{1152}  \langle \bar{q} g_s \sigma G q \rangle \nonumber \\
 &-& ig_s \frac{G^{\alpha \beta}_{ab}}{32 \pi^2 x^2} [\slashed x \sigma_{\alpha \beta}+ \sigma_{\alpha \beta} \slashed x]- i\delta_{ab} \frac{x^2 \slashed x g_s^2 \langle \bar{q}q \rangle^2}{7776} 
 - \delta_{ab} x^4 \frac{\langle \bar{q}q \rangle \langle g_s^2 G^2 \rangle}{27648} + \cdots.  \label{light}
\end{eqnarray}
The heavy quark propagator $S^{ab}_c(x)$ can be written as
\begin{eqnarray}
S^{ab}_c(x) &=& i \int \frac{d^4k}{(2\pi)^4} e^{-ikx} [  \frac{\delta_{ab} (\slashed k + m_c)}{k^2-m_c^2}  
- \frac{g_sG^{\alpha \beta}_{ab}}{4}\frac{\sigma_{\alpha \beta} (\slashed k + m_c) +(\slashed k + m_c)\sigma_{\alpha \beta}}{(k^2-m_c^2)^2} \nonumber \\
&+& \frac{g_s^2G^2 }{12} \delta_{ab} m_c \frac{k^2 +m_c \slashed k}{(k^2-m_c^2)^4} + \cdots
]. \label{heavy}
\end{eqnarray} 
Here $a,b=1,2,3$ and $A,B,C=1,2,\cdots,8$ are color indices.  $t^A$ is defined as $t^A=\lambda^A/2$, where $\lambda^A$ are the Gell-Mann matrices. $G^{A}$ which is the gluon field strength tensor, defined as $G^{A}_{\alpha \beta}=G^{A}_{\alpha \beta}(0)$ and is fixed at $x=0$.

The correlation function $\Pi^{\text{OPE}}(q)$ can be written by a dispersion integral
\begin{equation}
\Pi^{\text{OPE}}(q)=\int_{(m_c+m_s)^2}^\infty \frac{\rho^{\text{OPE}}(s)}{s-q^2},
\end{equation}
where we parametrize $s_{min}=(m_c+m_s)^2$. The  mass and decay constant (residue) sum rules can be obtained as mentioned before by equating $\Pi^{\text{Phys}}(q)$ and $\Pi^{\text{OPE}}(q)$. The mass can be found as
\begin{equation}
m_{X_0}^2=\frac{\int_{(m_c+m_s)^2}^{s_0}ds s \rho^{\text{OPE}}(s)e^{-s/M^2} }{\int_{(m_c+m_s)^2}^{s_0}ds  \rho^{\text{OPE}}(s)e^{-s/M^2} },
\end{equation}
whereas for decay constant $\lambda_{X_0}$ we use the formula
\begin{equation}
\lambda_{X_0}^2 m_{X_0}^2  e^{-m_{{X_0}}^2/M^2}=\int_{(m_c+m_s)^2}^{s_0}ds s \rho^{\text{OPE}}(s)e^{-s/M^2}.
\end{equation}
The physical properties can be extracted from the sum rules which depend on vacuum condensates and quark masses. We use the following values of quark masses and condensates \cite{Zyla:2020zbs}:
\begin{eqnarray}
m_c&=& (1.275^{+0.025}_{-0.035}) ~ \text{GeV}, \nonumber \\
m_s(2 ~ \text{GeV})&=& 95^{+9}_{-3 } ~ \text{MeV},\nonumber \\
m_0^2&=& (0.8 \pm 0.2) ~ \text{GeV}^2, \nonumber \\
\langle \bar{s}g_s \sigma G s \rangle &=& m_0^2 \langle \bar{s} s \rangle, \nonumber \\
\langle \bar{q}q \rangle &=& (-0.24 \pm 00.1)^3 ~ \text{GeV}^3, \nonumber \\
\langle \bar{s} s \rangle &=&(0.8 \pm 0.1)\langle \bar{q}q \rangle, \nonumber \\ 
\langle g^2G^2 \rangle &=& (0.012 \pm 0.004) ~ \text{GeV}^4. \nonumber  \\ \label{qcdinput}
\end{eqnarray}

The resulting sum rules for mass and decay constant is a function of Borel parameter $M^2$ and continuum threshold $s_0$. The correct choice of these parameters is an important task for sum rule calculations. One needs to find working regions for Borel and continuum threshold parameters where physical quantities do not extremely depend on these values or have a weak dependence on them. The QCDSR method have unavoidable theoretical uncertainties due to the its formulation. There exist some procedures to extract $M^2$ and $s_0$ values which are well defined in the context of the QCDSR method itself. In order to maintain desired results, working regions for $M^2$ and $s_0$ should satisfy some constraints on the pole contribution (PC)
\begin{equation}
\text{PC}= \frac{\Pi(M^2, s_0)}{\Pi(M^2, \infty)},
\end{equation} 
and OPE convergence
\begin{equation}
R(M^2)=\frac{\Pi^{\text{DimN}}(M^2,s_0)}{\Pi(M^2, s_0)}.
\end{equation}
Here, $\Pi^{\text{DimN}}(M^2,s_0)$ is a last term in the correlation function. We used quark, gluon, and mixing vacuum condensates up to dimension eight.

The continuum threshold $s_0$ is related to the energy of the first excited state of the studying hadron. If this first excited state exist then $s_0$ can be parametrized since the energy of this excited state is known. In the case of exotic resonances or states, it is not always possible to observe excited levels. Therefore it is difficult to determine $s_0$ precisely. However, some recipes exist to determine this energy level for exotic resonances and states.  

In standard recipes, the continuum threshold $s_0$ is taken to be as $s_0 = (m_{ground} + \delta)^2$ where $\delta$ varies between 0.3 and 0.8 GeV. Here we consider the LHCb Collaboration resonance as a ground state. The value of $s_0$ is determined from condition that the sum rule with 10 \% accuracy reproduce the experimentally measured mass. The working window for Borel parameter $M^2$ can be found by using PC and $R(M^2)$. PC is used to fix upper bound of $M^2$ and $R(M^2)$ can be used to find lower bound of $M^2$. In QCDSR formalism, the pole contribution  of the ground state is required to be greater than the contribution of the continuum states. OPE convergence is not guaranteed for $M^2 > 3 ~ \text{GeV}^2$ in the present work. Our analysis shows that for $\text{PC} > 0.5$ and $\text{R} < 0.2$, the working regions are
\begin{equation}
M^2 \in [2,3] ~ \text{GeV}^2, ~ s_0 \in [10,12] ~ \text{GeV}^2.
\end{equation}

\begin{figure}[H]
\centering
\includegraphics[width=3.5in]{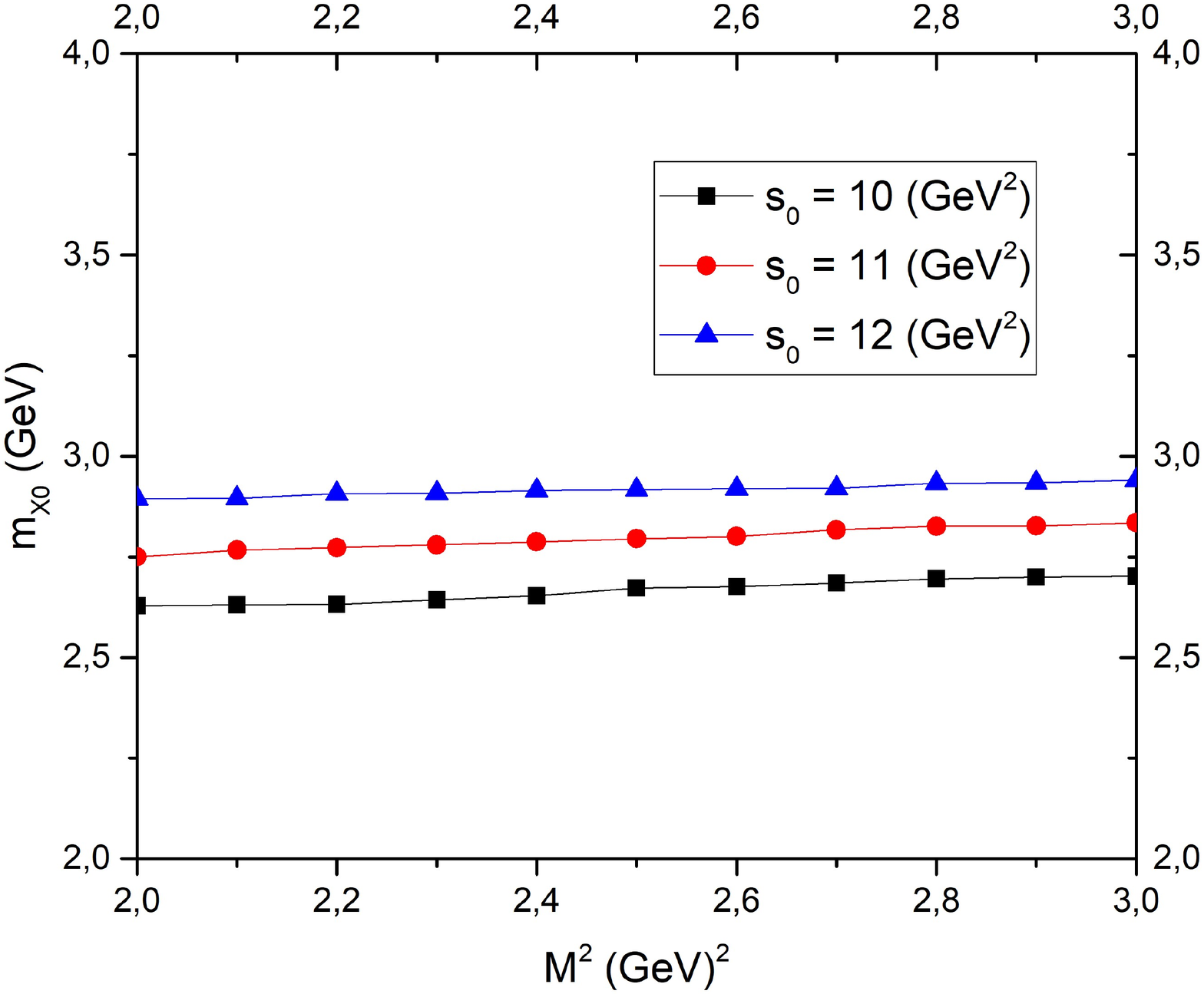}
\caption{\label{fig:fig1} The mass of the $X_0$ as a function of the Borel parameter $M^2$ at fixed $s_0$ values.} 
\end{figure}

\begin{figure}[H]
\centering
\includegraphics[width=3.5in]{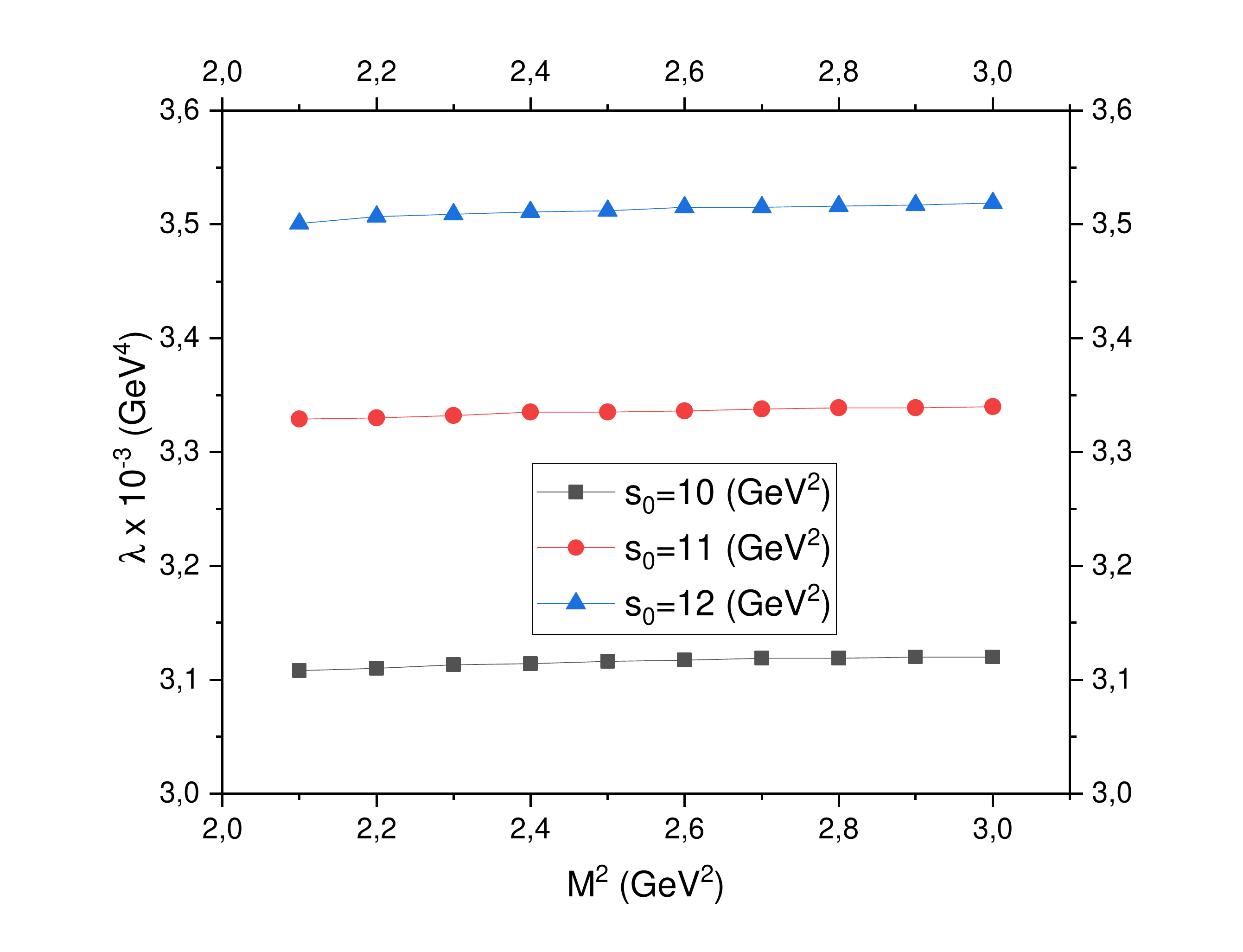}
\caption{\label{fig:fig2} The decay constant of the $X_0$ as a function of the Borel parameter $M^2$ at fixed $s_0$ values.} 
\end{figure}

The predicted value for the mass of $X_0$ is
\begin{equation}
M_{X_0}=2792 \pm 124  ~ \text{MeV},
\end{equation}
and decay constant  is
\begin{equation}
\lambda= 3.35 \pm 0.15   \times 10^{-3} ~ \text{GeV}^4.
\end{equation}
where the uncertainty results from the Borel mass parameter $M^2$, continuum threshold $s_0$, and various quark and gluon parameters. The extracted mass agree well with the mass of experimental result for $X_0(2900)$. In Figs. \ref{fig:fig1} and \ref{fig:fig2}, we display prediction of sum rules for mass $m$ and decay constant $\lambda$  as a function of Borel parameter $M^2$, respectively.

It can be seen from both figures that in the given working regions, there exist a mild dependence on the Borel parameter $M^2$ for the mass and decay constant results.

\begin{figure}[H]
\centering
\includegraphics[width=3.5in]{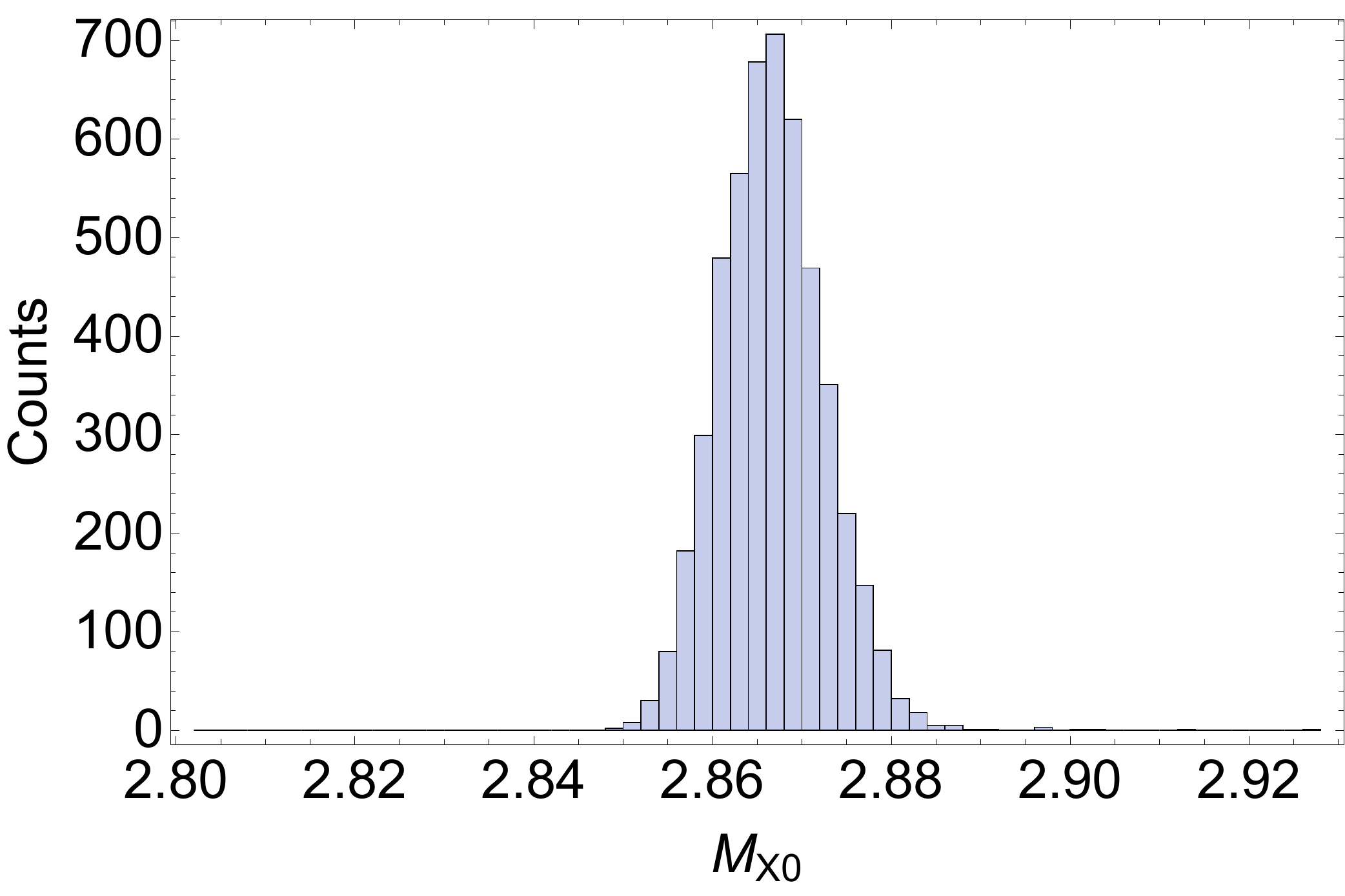}
\caption{\label{fig:fig3} The histogram of the $X_0$ masses obtained from 3000 matches.} 
\end{figure}

The next task is to study mass and decay constant of $X_0(2900)$ by using QCD sum rules with Monte-Carlo analysis. We set $n_B=50$ and generated 3000  Gaussian distributed input parameters with given uncertainties (10\% uncertainties, which are typical uncertainties in QCDSR). The QCD input parameters are quark condensate, gluon condensate, mixed condensate values which are given in Eq. \ref{qcdinput} and $\Lambda_{\text{QCD}}=0.353 ~ \text{GeV} $. We have selected the physical results since in the sum rules some constraints exist such as $s_0>m^2$. We also plot the histogram for 3000 different $X_0(2900)$ matches obtained in the least-squares fitting procedure. Fig. \ref{fig:fig3} shows the distribution of $X_{0}$ meson masses.

The distribution corresponds to
\begin{equation}
M_{X_0}=2865^{+20}_{-18} ~ \text{MeV},
\end{equation}
which is in good agreement with the experimental value. The uncertainty is lower than the commonly assumed 10 \%. It can be seen from Fig. \ref{fig:fig3} that the distribution of mass is very close to Gaussian curve. 

\begin{figure}[H]
\centering
\includegraphics[width=3.5in]{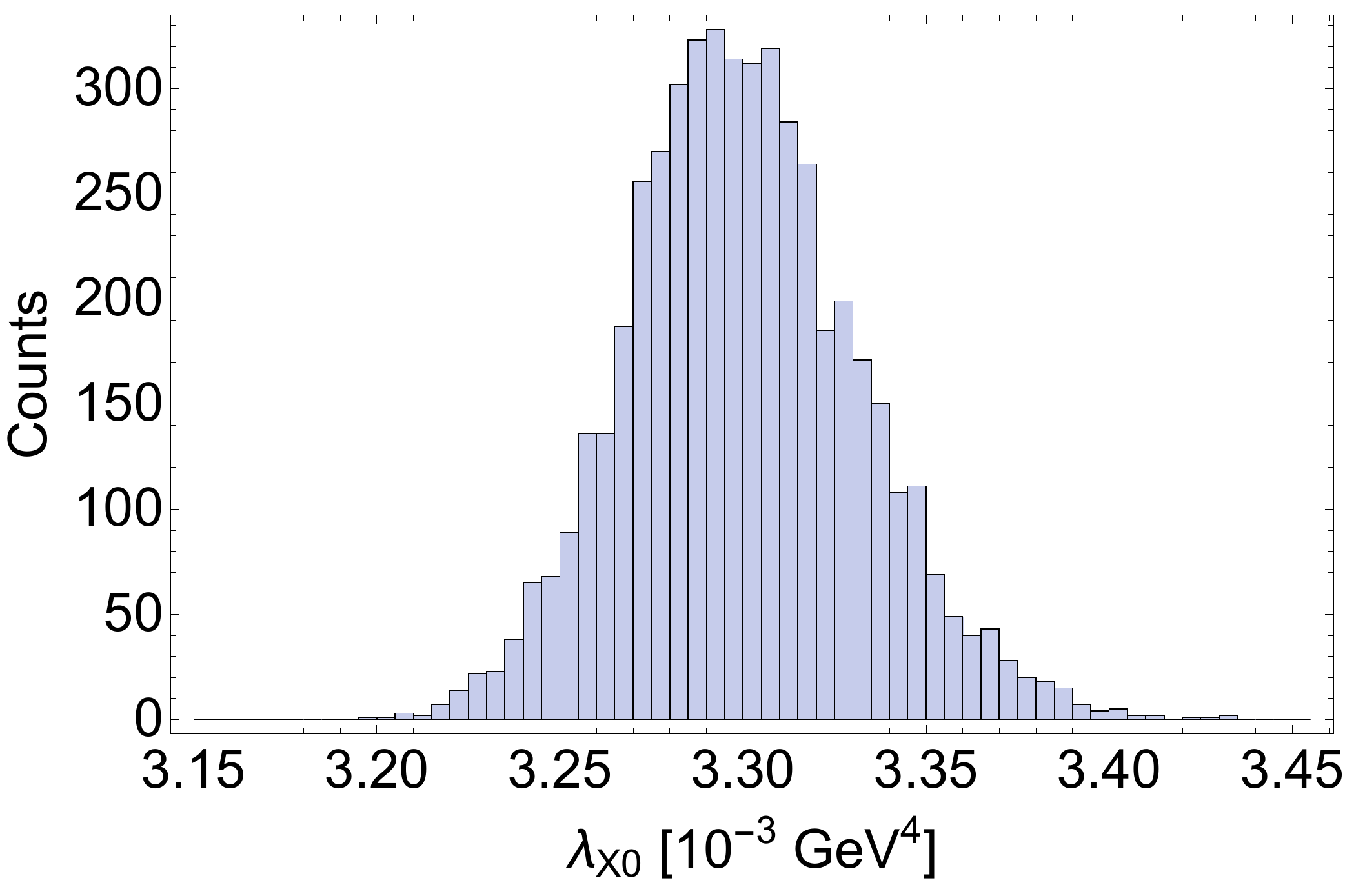}
\caption{\label{fig:fig4} The histogram of the $X_0$ decay constant obtained from 3000 matches.} 
\end{figure}

Fig. \ref{fig:fig4} shows the distribution of $X_{0}$ decay constants. 

The distribution corresponds to
\begin{equation}
\lambda_{X_0}=3.30^{+0.12}_{-0.14} ~ \times 10^{-3} ~ \text{GeV}^4.
\end{equation}
It can be seen from Fig. \ref{fig:fig4} that the distribution of decay constant is very close to Gaussian curve.

\section{\label{sec:level4}Traditional and Monte-Carlo Analysis of $X_1(2900)$ with $J^P=1^-$}
In this section, we investigate the diquark-antidiquark structure for the observed peak $X_1(2900)$ with quantum number $J^P=1^-$. We use the following interpolating current 
\begin{equation}
j_{\mu}(x)=s_a^T(x)Cc_b(x) \left[ \bar{u}_a(x) \gamma_{\mu} \gamma_5 C \bar{d}_b^T(x)- \bar{u}_b(x) \gamma_{\mu} \gamma_5  C \bar{d}_a^T(x) \right],
\end{equation}
where $a$ and $b$ are color indices and $C$ is the charge conjugation matrix. 

The correlation function can be written as
\begin{eqnarray}
\Pi_{\mu \nu}(q)&=&i \int d^4x e^{iqx} \langle 0 | T[j_{\mu}(x)j_{\nu}^{\dagger}(0)]| 0\rangle \nonumber \\ 
&=& \left( -g_{\mu \nu}+\frac{q_{\mu} q_{\nu}}{q^2}\right) \Pi_1(q) +  \frac{q_{\mu} q_{\nu}}{q^2} \Pi_0(q)\label{corx1},
\end{eqnarray}
where $\Pi_1(q)$  and $\Pi_0(q)$ are the invariant functions related to the spin-1 and spin-0 intermediate states, respectively. Spin-1 intermediate states contribute only to  $\Pi_1(q)$, therefore we use it to perform numerical analysis. In order to obtain sum rules expression, at first step correlation function needs to be calculated in terms of the physical parameters of the related hadron. Inserting a complete set of the $X_1(2900)$ state into the correlation function, we find
\begin{equation}
\Pi_{\mu \nu}^{\text{Phys}}(q)=\frac{\langle 0 \vert j_{\mu} \vert X_1(q) \rangle \langle X_1(q)\vert  j_\nu^{\dagger} \vert 0 \rangle}{m_{X_1}^2-q^2}+ \cdots,
\end{equation}
where the dots represents contributions from the higher resonances and continuum states. We can write coupling $\lambda_{X_1}$ using the matrix element 
\begin{equation}
\langle 0 \vert j_{\mu} \vert X_1(q) \rangle =\lambda_{X_1} m_{X_1} \epsilon_{\mu},
\end{equation}
where $\epsilon_{\mu}$ is the polarization vector of the $X_1(2900)$ state. With this definition correlation function can be written as
\begin{equation}
\Pi_{\mu \nu}^{\text{Phys}}(q)=\frac{\lambda_{X_1}^2 m_{X_1}^2}{m_{X_1}^2-q^2} \left(-g_{\mu \nu}+ \frac{q_{\mu}q_{\nu}}{m_{X_1}^2} \right)+ \cdots. \label{above}
\end{equation}
Applying Borel transformation to Eq. (\ref{above}) yields
\begin{equation}
\mathcal{B}_{q^2}  \Pi_{\mu \nu}^{\text{Phys}}(q)= \lambda_{X_1}^2  m_{X_1}^2  e^{-m_{X_1}^2/M^2}\left(-g_{\mu \nu}+ \frac{q_{\mu}q_{\nu}}{m_{X_1}^2} \right)+ \cdots
\end{equation}

At second step, the correlation function has to be calculated from the QCD (OPE) side. Contracting the heavy and light quark fields yields
\begin{eqnarray}
\Pi_{\mu \nu}^{\text{OPE}}(q)&=&i \int d^4x e^{iqx} \text{Tr} [\gamma_5 \tilde{S}^{aa^\prime}_s(x)\gamma_5 S_c^{bb^\prime}(x)] \{ \text{Tr} [\gamma_{\mu} \tilde{S}^{a^\prime b}_d(-x)\gamma_{\nu} S_u^{b^\prime a}(-x)] - \text{Tr} [\gamma_{\mu} \tilde{S}^{b^\prime b}_d(-x)\gamma_{\nu} S_u^{a^\prime a}(-x)] \nonumber \\ &+&  \text{Tr} [\gamma_{\mu} \tilde{S}^{b^\prime a}_d(-x)\gamma_{\nu} S_u^{a^\prime b}(-x)]- \text{Tr} [\gamma_{\mu} \tilde{S}^{a^\prime a}_d(-x)\gamma_{\nu} S_u^{b^\prime b}(-x)] \}.
\end{eqnarray}
Here we use the notation $\tilde{S}_q(x)=CS^T(x)C$, where $C$ is the charge conjugation operator. The spectral densities are given in the appendix. The $\Pi_{\mu \nu}^{\text{OPE}}(q)$ can be written by a dispersion integral
\begin{equation}
\Pi_{\mu \nu}^{\text{OPE}}(q)=\int_{(m_c+m_s)^2}^\infty ds \frac{\rho^{\text{OPE}}(s)}{s-q^2}.
\end{equation}
After normal procedure as defined previous section, the mass can be found as
\begin{equation}
m_{X_1}^2=\frac{\int_{(m_c+m_s)^2}^{s_0}ds s \rho^{\text{OPE}}(s)e^{-s/M^2} }{\int_{(m_c+m_s)^2}^{s_0}ds  \rho^{\text{OPE}}(s)e^{-s/M^2} },
\end{equation}
whereas for the decay constant $\lambda_{X_1}$ we use the formula
\begin{equation}
\lambda_{X_1}^2 m_{X_1}^2  e^{-m_{{X_1}}^2/M^2}=\int_{(m_c+m_s)^2}^{s_0}ds s \rho^{\text{OPE}}(s)e^{-s/M^2}.
\end{equation}
The working regions for continuum threshold $s_0$ and Borel parameter $M^2$ are the same as given in previous section. The predicted value for the mass of $X_1$ is
\begin{equation}
M_{X_1}=2963 \pm 64  ~ \text{MeV}
\end{equation}
which is in good agreement with the mass of experimental result for $X_1(2900)$. The predicted decay constant of $X_1(2900)$ is
\begin{equation}
\lambda= 3.39 \pm 0.20   \times 10^{-3} ~ \text{GeV}^4.
\end{equation}

\begin{figure}[H]
\centering
\includegraphics[width=3.5in]{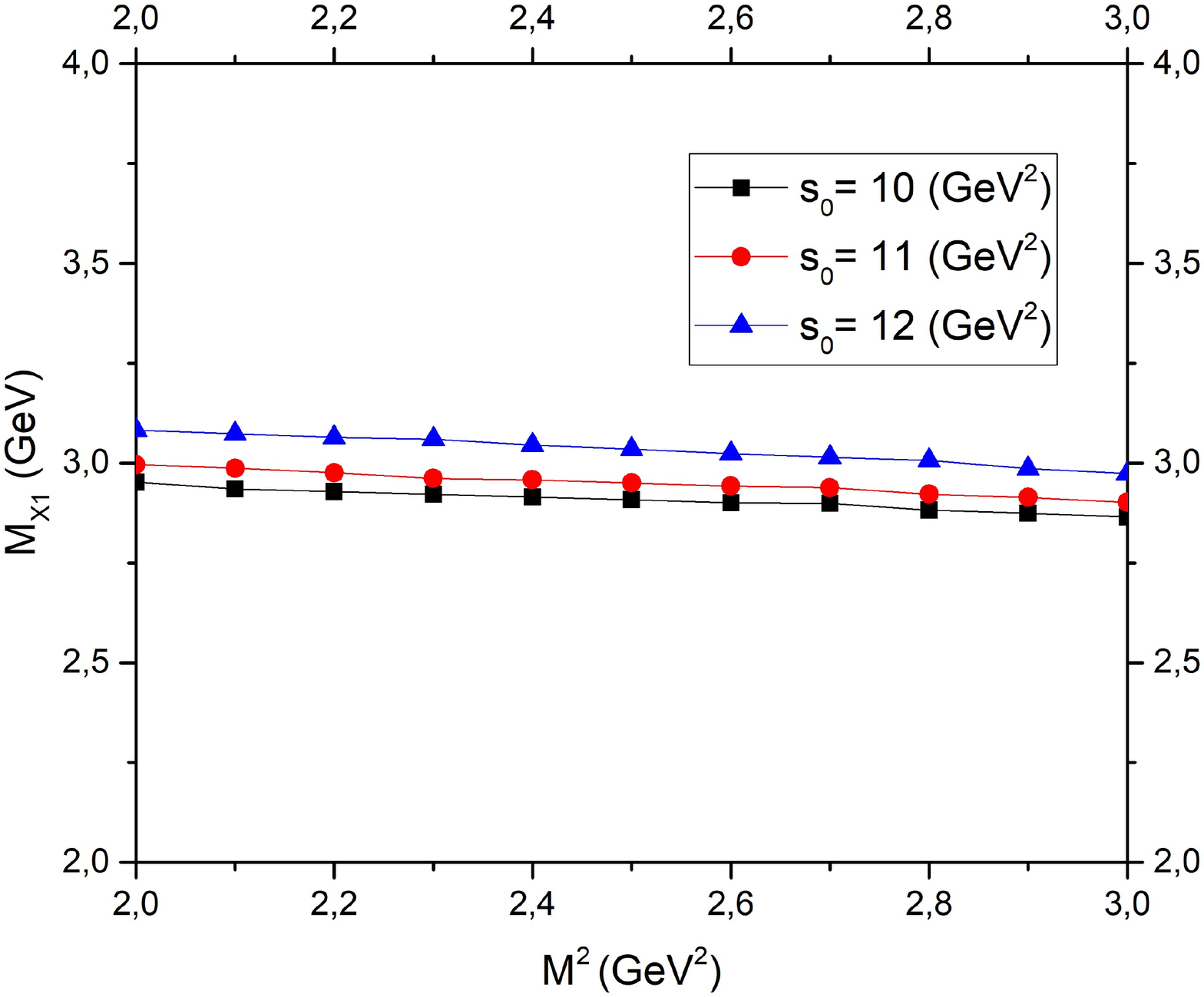}
\caption{\label{fig:fig5} The mass of the $X_1$ as a function of the Borel parameter $M^2$ at fixed $s_0$ values.} 
\end{figure}

\begin{figure}[H]
\centering
\includegraphics[width=3.5in]{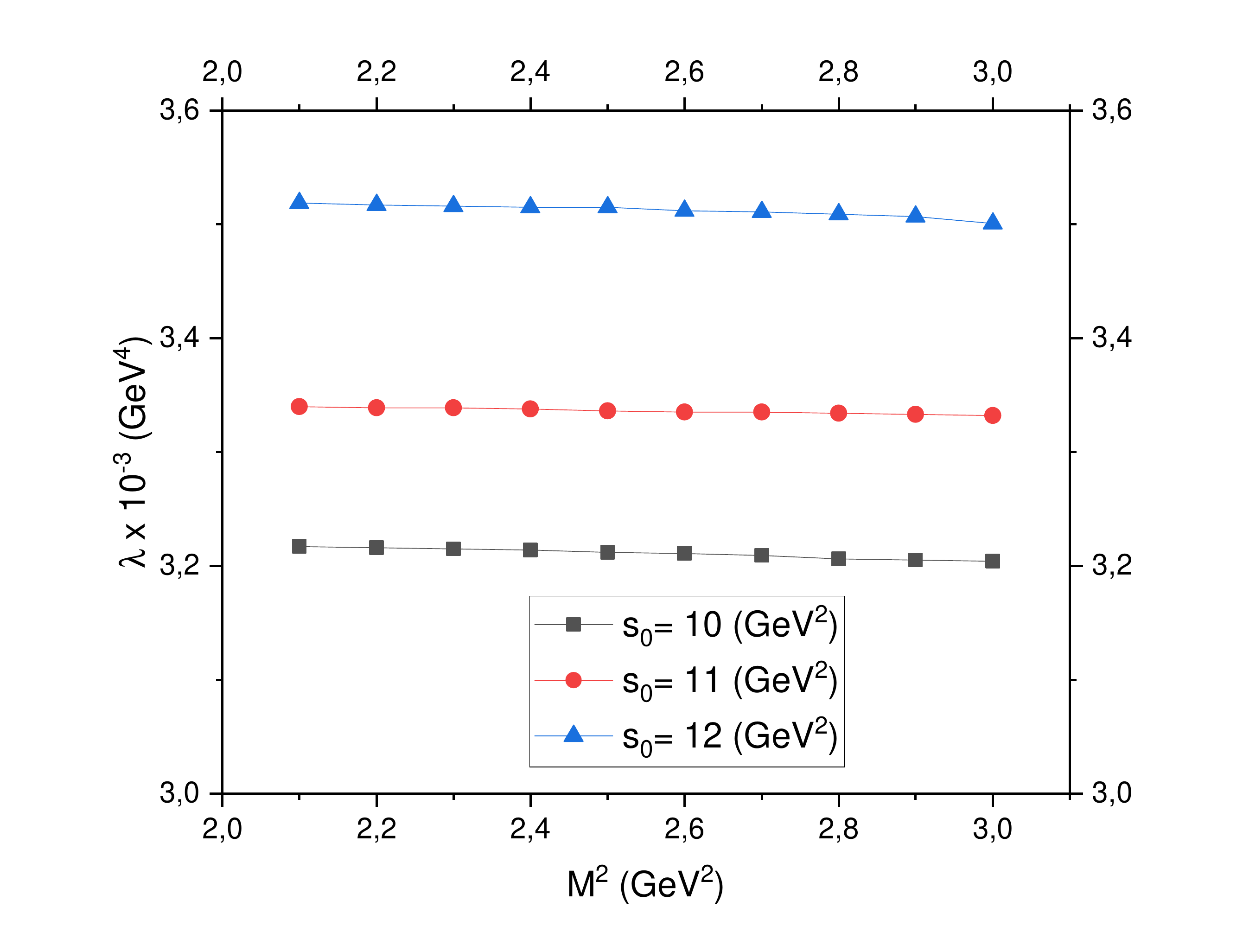}
\caption{\label{fig:fig6} The decay constant of the $X_1$ as a function of the Borel parameter $M^2$ at fixed $s_0$ values.} 
\end{figure}

In Figs. \ref{fig:fig5} and \ref{fig:fig6} we display prediction of sum rules for mass $m$ and decay constant $\lambda$ as a function of Borel parameter $M^2$, respectively.

It can be seen from Figs. \ref{fig:fig5} and \ref{fig:fig6} that  the predictions of sum rule for mass $m$ and decay constant $\lambda$ as a function of Borel parameter $M^2$ have a smooth dependence on the Borel parameter. 

\begin{figure}[H]
\centering
\includegraphics[width=3.5in]{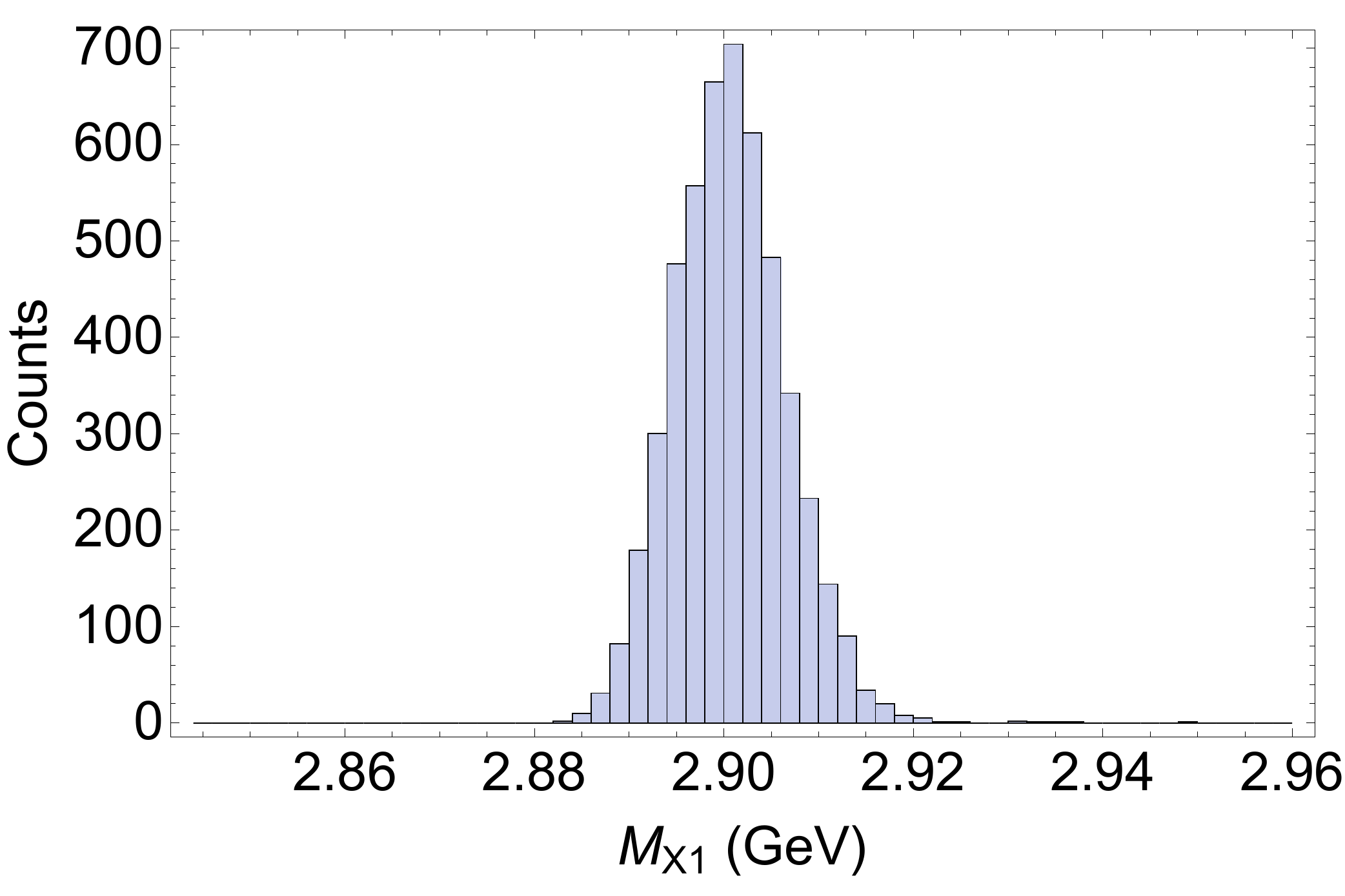}
\caption{\label{fig:fig7} The histogram of the $X_1$ masses obtained from 3000 matches.} 
\end{figure}

In order to obtain mass and decay constant of $X_1(2900)$ we set $n_B=50$ and generated 3000  Gaussian distributed input parameters with given uncertainties (10\% uncertainties, which are typical uncertainties in QCDSR). The QCD input parameters are quark condensate, gluon condensate, mixed condensate values which are given in Eq. \ref{qcdinput} and $\Lambda_{\text{QCD}}=0.353 ~ \text{GeV} $. We have selected the physical results since in the sum rules some constraints exist such as $s_0>m^2$. We also plot the histogram for 3000 different $X_0(2900)$ matches obtained in the least-squares fitting procedure. The histogram for masses of $X_1$ can be seen in Fig. \ref{fig:fig7}. 

\begin{figure}[H]
\centering
\includegraphics[width=3.5in]{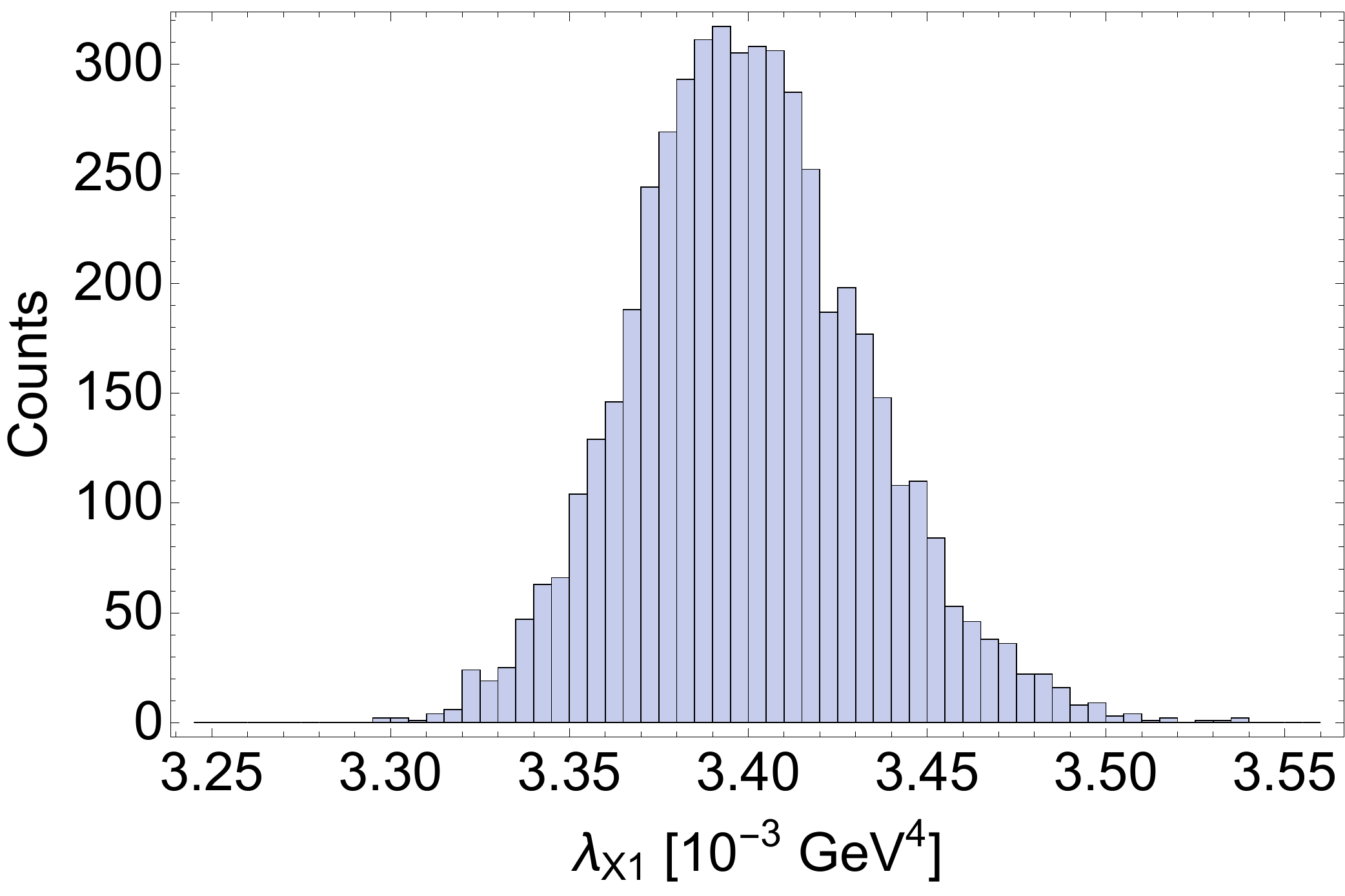}
\caption{\label{fig:fig8} The histogram of the $X_1$ decay constants obtained from 3000 matches.} 
\end{figure}

The distribution corresponds to 
\begin{equation}
M_{X_1}=2900 \pm 19 ~ \text{MeV},
\end{equation}

which is in good agreement with the experimental value. The uncertainty is lower than the commonly assumed 10 \%. The distribution of mass is very close to Gaussian curve. Fig. \ref{fig:fig8} shows the distribution of decay constants $\lambda$ for the $X_1(2900)$.

The distribution corresponds to
\begin{equation}
\lambda_{X_1}=3.40^{+0.15}_{-0.13} ~ \times 10^{-3} ~ \text{GeV}^4.
\end{equation}
It can be seen from Fig. \ref{fig:fig8} that the distribution of decay constant is very close to Gaussian curve.

\section{\label{sec:level5}Discussion, Summary and Concluding Remarks}
In this present work, we have studied two new resonances $X_0$ and $X_1$ observed by the LHCb collaboration. This is the first time observing fully open-flavor exotic states. We studied their possible interpretations using traditional and Monte-Carlo based QCD sum rules method. 

We used a molecular current (meson-meson tetraquark) with $J^P=0^+$ to study mass and decay constant of $X_0(2900)$. In the traditional analysis, the extracted mass value is 
\begin{equation}
M_{X_0}=2792 \pm 124  ~ \text{MeV},
\end{equation}
and in good agreement with the experimental mass of $X_0(2900)$  suggesting a possible molecular picture as $J^P=0^+$ $D^{-}K^{+}$. This molecular picture was supported in \cite{Molina:2010tx}, in which the authors made a prediction for the $0^+$ state with mass $2848 ~  \text{MeV}$ and $\Gamma = 59 ~  \text{MeV}$ before the experiment. The decay constant of $X_0(2900)$ is obtained as 
\begin{equation}
\lambda= 3.35 \pm 0.15   \times 10^{-3} ~ \text{GeV}^4.
\end{equation}

The Monte-Carlo analysis for the mass and decay constant of $X_0(2900)$ yields mass
as
 \begin{equation}
M_{X_0}=2865^{+20}_{-18} ~ \text{MeV},
\end{equation}
and decay constant as
\begin{equation}
\lambda_{X_0}=3.30^{+0.12}_{-0.14} ~ \times 10^{-3} ~ \text{GeV}^4.
\end{equation}
Our analysis give results which are in good agreement with the Monte-Carlo analysis. 

A diquark-antidiquark tetraquark interpolating current with $J^P=1^-$ was used to study mass and decay constant of $X_1(2900)$. The extracted mass value is 
\begin{equation}
M_{X_1}=2963 \pm 64  ~ \text{MeV}
\end{equation}
which is in good agreement with the experimental mass of $X_1(2900)$. The obtained decay constant  is
\begin{equation}
\lambda= 3.39 \pm 0.20   \times 10^{-3} ~ \text{GeV}^4.
\end{equation}

The Monte-Carlo analysis for the mass and decay constant of $X_1(2900)$ yields mass
as
 \begin{equation}
M_{X_0}=2900 \pm 19 ~ \text{MeV},
\end{equation}
and decay constant as
\begin{equation}
\lambda_{X_1}=3.40^{+0.15}_{-0.13} ~ \times 10^{-3} ~ \text{GeV}^4.
\end{equation}
Our analysis give results which are in good agreement with the Monte-Carlo analysis. Comparison of mass results are tabulated in \ref{tab:table1} whereas decay constant results are tabulated in \ref{tab:table2}

\begin{table}[H]
\caption{\label{tab:table1} Mass values of traditional sum rule analysis vs. Monte-Carlo analysis. All results are in MeV.}
\begin{ruledtabular}
\begin{tabular}{cccc}
  & Traditional & Monte-Carlo  & Experiment
  \\
\hline
$X_0$ & $2792 \pm 124$ & $2865^{+20}_{-18} $ &$2866 \pm 7$    \\
$X_1$  & $2963 \pm 64$ & $2900 \pm 19 $ & $2904 \pm 5$  \\
\end{tabular}
\end{ruledtabular}
\end{table}

\begin{table}[H]
\caption{\label{tab:table2} Decay constants of traditional sum rule analysis vs. Monte-Carlo analysis.}
\begin{ruledtabular}
\begin{tabular}{cccc}
  & Traditional & Monte-Carlo 
  \\
\hline
$X_0$ & $3.35 \pm 0.15   \times 10^{-3} ~ \text{GeV}^4$ & $3.30^{+0.12}_{-0.14} ~ \times 10^{-3} ~ \text{GeV}^4$    \\
$X_1$  & $3.39 \pm 0.20   \times 10^{-3} ~ \text{GeV}^4$ & $3.40^{+0.15}_{-0.13} ~ \times 10^{-3} ~ \text{GeV}^4$  \\
\end{tabular}
\end{ruledtabular}
\end{table}

As can be seen from both tables, results for the both of the traditional analysis and Monte-Carlo analysis  are in good agreement with experimental values.

An interesting feature of this experiment is that $X_1(2900)$ has larger width than the $X_0(2900)$. For this purpose, we made a Monte-Carlo based analysis for the masses of these states. It can bee seen form Figs. \ref{fig:fig3} and \ref{fig:fig7}, $X_0(2900)$ has a narrower shape than the $X_1(2900)$. In other words, $X_0(2900)$ has a small width than the $X_1(2900)$ has. This analysis could support the molecular picture for $X_0(2900)$ and diquark-antidiquark tetraquark structure for $X_1(2900)$. This outcome could be tested by calculating decay and width properties. For example, the width of the $X_0(2900)$ state in molecular picture was calculated in \cite{Agaev:2020nrc}. The result is $\Gamma=(49.6 \pm 9.3)~ \text{MeV}$ and could be considered as to favor molecular picture. In \cite{Lu:2020qmp}, it is found that a compact tetraquark assignment for $X_0(2900)$ is not likely.  A QCD sum rule study suggested the tetraquark configuration for $X_0(2900)$ \cite{Zhang:2020oze}. It is pointed out that $X_1(2900)$ has a width around $\Gamma \sim 100~ \text{MeV}$  and could be interpreted as P-wave compact tetraquark \cite{Chen:2020aos}. It was showed in \cite{He:2020jna} that $X_0(2900)$ and $X_1(2900)$ states ban be explained as radial  excited tetraquark and orbitally excited tetraquark of $[ud\bar{s}\bar{c}]$ configurations, respectively. 

In QCD sum rule method, one cannot conclude if a state have a tetraquark  or molecular configuration. However, both a traditional and Monte-Carlo analysis with mass and decay constant could support the possible assignments. The two interpretations made above are just possible assignments for these states. More theoretical and experimental studies are needed to investigate these exotic states.  

\appendix
\section{Spectral Density of $X_0(2900)$ Resonance}
With the definitions
\begin{eqnarray}
p &=& [m_c^2+s(x-1)], \nonumber \\
q &=& [m_c^2+2s(x-1)], \nonumber \\
r &=& [m_c^2+3s(x-1)], \nonumber \\
s &=& [m_c^2+5s(x-1)], \nonumber \\
t &=& m_c m_s \langle s\bar{s} \rangle (1-x), \nonumber 
\end{eqnarray}

we give the spectral density expressions of correlation functions for the $X_0(2900)$ resonance.

\begin{eqnarray}
\rho^{\text{Pert}}&=&\frac{1}{2^{13} \pi^6} \int_{(m_c^2+m_s^2)}^{s_0}\frac{dx x^4}{(1-x)^3}p^3r, \nonumber \\
\rho^{3}&=&\frac{3}{2^8 \pi^4} \int_{(m_c^2+m_s^2)}^{s_0}\frac{dx x^2}{(1-x)^2}p \left[  t(\frac{m_c^2+1}{m_c}) -m_c^2 \langle q\bar{q} \rangle -2m_s \langle q\bar{q} \rangle (1-x)(r-s(x-1))-2m_s s  \langle s\bar{s} \rangle (x-1)^2 \right], \nonumber \\ 
\rho^{4}&=&\frac{1}{3 2^6 \pi^4} \langle \frac{\alpha_s G^2}{\pi} \rangle\int_{(m_c^2+m_s^2)}^{s_0}\frac{dx x^2}{(1-x)^3} \left[ 2m_c^4 (x(13x-30)+18)+3m_c^2s (6-5x)^2(x-1)+24s^2(x-1)^3(2x-3)\right], \nonumber \\
\rho^{5}&=&\frac{m_0^2}{ 2^8 \pi^4} \int_{(m_c^2+m_s^2)}^{s_0}\frac{dx x}{(1-x)} \left[ -3m_c^3 \langle q\bar{q} \rangle +2m_c^2 m_s \langle s\bar{s}\rangle (1-x) +3m_c  s \langle q\bar{q} \rangle (1-x) - 3m_s (1-x) (\langle q\bar{q} \rangle - s \langle s\bar{s} \rangle (1-x))\right], \nonumber \\
\rho^{6}&=&\frac{1}{ 2^6 \pi^4} \int_{(m_c^2+m_s^2)}^{s_0}dx x \left[ \langle g^3 G^3 \rangle \frac{x^4}{2^5 5 \pi^2 (1-x)^3} (m_c^2 (2x+3)+s (5x^2-7x+2))-g_s^2/3^3 (m_c^2+r) (2\langle q\bar{q} \rangle ^2+\langle s\bar{s} \rangle)\right], \nonumber \\
\rho^{7}&=&\frac{1}{3 2^4 \pi^2}  \langle \frac{\alpha_s G^2}{\pi} \rangle  \int_{(m_c^2+m_s^2)}^{s_0} \frac{dx}{(1-x)^2} \left[ m_c^2 \langle q\bar{q} \rangle (2 x^3+7x^2-14x+7)+2m_s \langle q\bar{q} \rangle  (5x^3-9x^2+3x +1)-6m_s \langle s\bar{s} \rangle  (x^3-2x^2+x) \right] \nonumber.
\end{eqnarray}

\section{Spectral Density of $X_1(2900)$ Resonance}
In this section, we give the spectral density expressions of correlation functions for the $X_1(2900)$ resonance.
\begin{eqnarray}
\rho^{\text{Pert}}&=&\frac{-1}{2^{13} \pi^6} \int_{(m_c^2+m_s^2)}^{s_0}\frac{dx x^4}{(1-x)^3}p^3s, \nonumber \\
\rho^{3}&=&\frac{1}{2^7 \pi^4} \int_{(m_c^2+m_s^2)}^{s_0}\frac{dx x^2}{(1-x)^2}p \left[ 2pm_c\langle q\bar{q} \rangle -rm_s(1-x) (\langle s\bar{s} \rangle-2\langle q\bar{q} \rangle)  \right], \nonumber \\
\rho^{4}&=&\frac{-1}{3^2 2^10 \pi^4} \langle \frac{\alpha_s G^2}{\pi} \rangle\int_{(m_c^2+m_s^2)}^{s_0}\frac{dx x^2}{(1-x)^3} \left[ m_c^4 (x(13x+21)-9)+2m_c^2s (x(23x-40+18)(x-1))-s^2 (1-x)^3 (32x-27)\right], \nonumber \\
\rho^{5}&=&\frac{m_0^2}{3 2^6 \pi^4} \int_{(m_c^2+m_s^2)}^{s_0}\frac{dx x}{(1-x)} \left[ 3m_c p \langle q\bar{q} \rangle  - m_s (1-x)q(\langle s\bar{s}\rangle -3 \langle q\bar{q} \rangle) \right], \nonumber \\
\rho^{6}&=&\frac{1}{3 5  2^12 \pi^6} \int_{(m_c^2+m_s^2)}^{s_0}dx x \left[ m_c^2 \langle g^3 G^3 \rangle x q + g_s^2  x^2 q  + 36 m_c m_s \langle q\bar{q} \rangle (\langle s\bar{s}\rangle -2 \langle q\bar{q} \rangle) \right], \nonumber \\
\rho^{7}&=&\frac{1}{3^2 2^7 \pi^2}  \langle \frac{\alpha_s G^2}{\pi} \rangle  \int_{(m_c^2+m_s^2)}^{s_0} \frac{dx}{(1-x)} \left[8 m_c^ \langle q\bar{q} \rangle + m_s (3x \langle s\bar{s} \rangle)+ 2 \langle q\bar{q} \rangle(1-4x) \right] \nonumber.
\end{eqnarray}

\end{document}